\documentclass[pmlr]{jmlr}

\usepackage{longtable}

\usepackage{booktabs}
\usepackage[load-configurations=version-1]{siunitx} 

\theorembodyfont{\upshape}
\theoremheaderfont{\scshape}
\theorempostheader{:}
\theoremsep{\newline}

\jmlrvolume{}
\jmlryear{2026}
\jmlrworkshop{Impactful and Responsible AI Systems for Education}

\title[Geometry Figure Annotation Tool]{A Two-Validator Web Interface for Structured Geometry Figure Annotation}

\author{\Name{Sabin-Codruț Badea} \Email{sabin-codrut.badea@s.unibuc.ro}\and
   \Name{Adrian-Marius Dumitran} \Email{marius.dumitran@unibuc.ro}\\
   \addr University of Bucharest, Faculty of Mathematics and Computer Science}

\editor{Editor's name}

\begin{document}

\maketitle

\begin{abstract}
Annotating geometric figures from scanned documents has long been addressed by adapting generic annotation tools, tools not originally designed for such tasks, to use cases where they are suboptimal. An interactive web interface is described that is purpose-built for validating automatically generated geometry figure descriptions, allowing annotators to review and correct conditional declaration language (CDL) descriptions while simultaneously adjusting figure crops and editing source problem text. Submissions pass through two independent annotators in sequence, with each round fully logged. The interface is currently deployed and has been used by 12 annotators to validate 483 problem entries.
\end{abstract}


\noindent\textbf{Code and Deployment:} Implementation and deployment guide available at \url{https://github.com/badea-codrut-cti/GeoValidator}\footnote{Repository with the platform source code.}.

\section{Introduction}
\label{sec:intro}

The digitization of olympiad-level geometric corpora is essential for building high-quality datasets that power intelligent tutoring systems, automated exercise generators, and adaptive hint engines for geometry education~\citep{chen-geometry3k, lu-mathvista}. OCR systems extract problem text with reasonable accuracy, and multimodal models can generate initial figure descriptions, but both stages produce noisy outputs that must be human-validated before they can safely be used in educational applications, because a single mis-labelled diagram or spoiled construction can invalidate an entire worked solution.

Generic annotation platforms (e.g., CVAT, Label Studio, Supervisely) offer robust project management but were not designed for structured mathematical content. Conversely, LaTeX editors such as Overleaf and Mathcha provide syntax validation but no image cropping or diagram annotation capabilities. No existing platform combines geometry-focused image cropping with structured description validation for geometric figures.

This gap matters because geometry reasoning datasets require precise figure-diagram annotations alongside problem text~\citep{chen-geometry3k, cao-geoqa}. In educational settings, rigorous validation of AI-generated descriptions acts as a lightweight governance layer for training and evaluation corpora, ensuring that downstream tools such as benchmark suites, automated solvers, or tutoring software are grounded in human-verified facts. Quality annotation frameworks like CrowdTruth~\citep{vecchi-crowdtruth} and Headwork~\citep{tu-headwork} address crowdsourcing quality but do not target mathematical content.

The present work addresses this gap by describing a lightweight, serverless web interface for validating structured geometry figure annotations, deployed on a free-tier edge platform to keep the infrastructure accessible to teachers and small research groups.

\section{System Design}
\label{sec:system}

\subsection{Interface Layout}

\begin{figure}[htbp]
  \centering
  \includegraphics[width=0.8\textwidth]{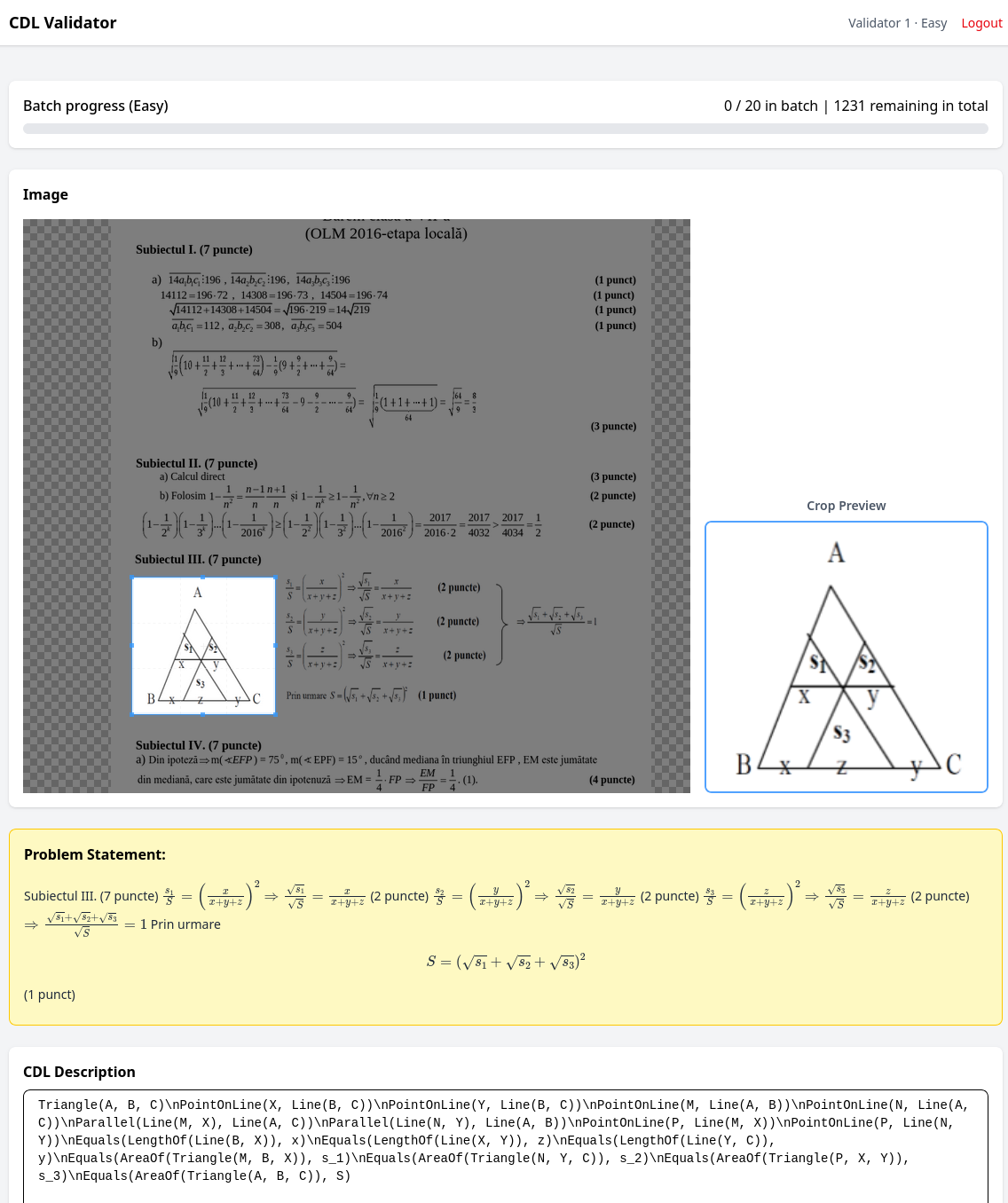}
  \caption{Screenshot of the three-panel validation interface.}
  \label{fig:interface}
\end{figure}

The interface uses a three-panel layout: problem statement with MathJax rendering (editable for OCR correction), source image with cropped diagram and Cropper.js re-cropping, and CDL description as plain-text editor (Figure \ref{fig:interface}).

\subsection{CDL Notation Development}
\label{sec:cdl-dev}

CDL was originally proposed by Zhang et al.~\citep{zhang-formalgeo} as the conditional declaration language within the FormalGeo framework, consisting of construction, condition, and goal statements. In that formal system, CDL is intentionally low-level so that every geometric property is stated atomically. For instance, a parallelogram must be declared through four separate \texttt{Parallel} predicates and two pairs of \texttt{Equals(LengthOf(...))} statements. While this granularity is essential for automated theorem provers, it becomes prohibitively verbose for multimodal LLMs, which must generate descriptions in a single forward pass and which rely on chain-of-thought reasoning that scales with output length. Higher token counts increase both inference cost and error rate, and the resulting descriptions are harder for human validators to parse.

We therefore deviate from the original FormalGeo CDL by introducing higher-level composite predicates that preserve semantic coverage but reduce statement count. Examples include \texttt{Parallelogram(A,B,C,D)} in place of four parallel-line declarations, and \texttt{Midpoint(M,Line(A,B))} in place of two collinearity and length-equality statements. These refinements cut average description length by roughly half while remaining expressible in FormalGeo's underlying predicate logic. Table~\ref{tab:cdl-predicates} summarizes the current predicate hierarchy.

\begin{table}[htbp]
  \centering
  \begin{tabular}{@{}lll@{}}
    \toprule
    Category & Purpose & Example predicates \\
    \midrule
    Shape declarations & 2D/3D primitives & \texttt{Triangle(A,B,C)} \\
    Structural relations & Geometric properties & \texttt{Parallel(Line(A,B),Line(C, D))}  \\
    Semantic measurements & Numeric constraints & \texttt{Equals(LengthOf(Line(A,B),6))} \\
    Edge cases & Arbitrary/unrepresentable & \texttt{Equals(..., Latex(\textbackslash sqrt[3]\{62\}))} \\
    \bottomrule
  \end{tabular}
  \caption{CDL predicate categories and representative examples.}
  \label{tab:cdl-predicates}
\end{table}

The predicate set was refined iteratively over 12 rounds, starting from a minimal subset and expanding based on observed annotation patterns. Of the 483 validated entries, 67 required no modifications to the AI-generated CDL output.

\begin{figure}[htbp]
  \centering
  \begin{minipage}{0.45\textwidth}
    \centering
    \includegraphics[width=\linewidth]{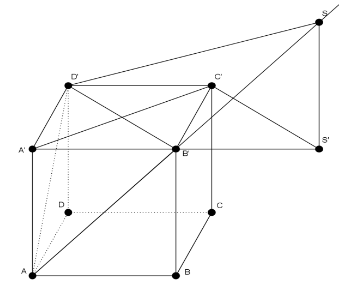}
  \end{minipage}
  \hfill
  \begin{minipage}{0.5\textwidth}
\begin{verbatim}
% 3D example
Cube(A, B, C, D, A', B', C', D')
Equals(LengthOf(Line(A, B)), 6)
PointOnLine(B', Line(A, S))
PointOnLine(B', Line(A', S'))
Perpendicular(Line(S, S'), Line(A', S'))
Equals(MeasureOf(Angle(D', S, B')), 30)
\end{verbatim}
  \end{minipage}
  \caption{Example of 3D CDL description for a cube geometry problem.}
  \label{fig:cdl-3d}
\end{figure}

\subsection{Two-Pass Validation Workflow}
Each entry is reviewed by two annotators sequentially. The first annotator corrects AI-generated CDL descriptions, adjusts figure crops, and edits problem text. Submissions without modifications are recorded as approvals; any edit stores the corrected content in per-annotator fields. The second annotator reviews the same entry with the first's corrections pre-filled, completing the entry. Every field has first-validator and second-validator columns, preserving original AI output alongside corrections.

\section{Limitations and Future Work}
\label{sec:limitations}

Validator agreement was substantial but not perfect. Of the completed entries, both annotators agreed on the final CDL, crop, and problem text in 67 cases (approximately one in seven). Agreement was determined on the basis that no changes were made to the crop, problem statement or CDL description, which was uncommon due to the fact that the first validator could potentially miss a tighter, more accurate crop of the geometric figure or a raw Unicode character in the problem statement that could be rewritten using inline Mathjax notation (i.e \texttt{\textbackslash cap} instead of $\cap$).

The majority of second-pass corrections to the CDL description were not disagreements over geometric facts, but rather refinements to prevent the description from \emph{spoiling} the problem, which was the most frequent failure mode we observed. For example, when a problem asks the student to \emph{prove} that a quadrilateral is a square, the model often described the figure as \texttt{Square(...)} because the drawing \emph{looks} square, even though the predicate itself encodes the conclusion to be proved. Maintaining pedagogically faithful descriptions is critical for educational downstream use: a tutoring system that pre-answers its own exercise, or a benchmark that supplies the theorem under test, undermines learning and invalidates evaluation. The two-pass workflow is deliberately designed to catch such didactic leakage.

Future work includes automated pre-validation steps that detect potential spoilers (e.g., flagging predicates whose consequences match the problem objective), formal inter-annotator agreement statistics on predicate-level disagreements, and extending the pipeline to other national olympiad corpora.

\section{Ethics Statement}
\label{sec:ethics}

All geometry problems used in this work were extracted from publicly available Romanian olympiad materials. The annotation platform records which session annotated which images but does not store any personally identifiable information about annotators. All annotators who participated in the validation process were adult volunteers and provided informed consent for the use of their annotation data.

\section{Conclusion}
\label{sec:conclusion}

We described a web interface for validating AI-generated geometry figure descriptions. The tool combines problem text editing, figure cropping, and CDL review in a two-pass workflow with full edit logging, and it runs on free serverless infrastructure.

\bibliography{references}

\end{document}